\documentclass[a4paper,longmktitle,authoryear]{cas-sc}

\usepackage[T1]{fontenc}
\usepackage[utf8]{inputenc}
\usepackage[english]{babel}

\usepackage[numbers]{natbib}
\usepackage{xpatch}
\usepackage{dsfont}
\usepackage{mathrsfs}
\usepackage{url}            
\usepackage{booktabs}       
\usepackage{amsfonts}       
\usepackage{nicefrac}       
\usepackage{microtype}      
\usepackage{lipsum}         %
\usepackage{graphicx}       %
\usepackage{color}          %
\usepackage{amsmath}
\usepackage{amssymb}
\usepackage{caption}
\usepackage{subcaption}
\usepackage{multirow}
\usepackage{multicol}
\usepackage{times}
\usepackage{graphicx}
\usepackage{breakcites}
\usepackage{natbib,enumitem}
\usepackage{bm} 
\usepackage{bbm}
\usepackage{epsfig}
\usepackage{alltt}
\usepackage{tikz}
\usepackage{caption}
\usepackage{booktabs}
\usepackage{pgfplots}
\pgfplotsset{compat=1.6}
\usepackage{xcolor}

\usepackage{algorithm}
\usepackage{algorithmic}
\usepackage[ruled,vlined, algo2e]{algorithm2e}
\usepackage{tcolorbox}
\usepackage{tikz}
\usepackage{tikz}
\usepackage{tikz-qtree,tikz-qtree-compat}

\begin{document}
\let\WriteBookmarks\relax
\def\floatpagepagefraction{1}
\def\textpagefraction{.001}
\shorttitle{}
\shortauthors{Victoire Djimna et~al.}

\title [mode = title]{Boosting the Predictive Accurary of Singer Identification Using Discrete Wavelet Transform For Feature Extraction}

\tnotetext[1]{This document is the results of the research
   project funded by AIMS CAMEROON with the help of MASTERCARD Foundation.}

\author[1]{Victoire Djimna Noyum}[orcid=0000-0002-0118-3668]
\cormark[1]
\ead{victoire.djimna@aims-cameroon.org}

\credit{Conceptualization of this study, Methodology, Software, Data curation, Writing - original draft}

\address[1]{School of Mathematical Sciences, African Institute for Mathematical Sciences, Crystal Garden, Limbe}

\author[1]{Younous  Perieukeu Mofenjou}[style=chinese]
\ead{younous.mofenjou@aims-cameroon.org}
\credit{Software, Result compilation, Writing -review}

\author[1]{Cyrille Feudjio}
\ead{cyrille.feudjio@aims-cameroon.org}
\credit{Software, Result compilation, Writing -review}

\author%
[2]
{Alkan  G\"{o}ktug}
\ead{alkang@aims-cameroon.org}

\credit{Supervision, Validation, Writing - review \& editing}

\address[2]{School of Mathematical Sciences, ETH Zurich, Rämistrasse 101, 8092 Zurich, Switzerland}

\author%
[3] 
{ Ernest Fokou\'e }
\ead{epfeqa@rit.edu}
\credit{Supervision, Software, Validation, Writing - review \& editing}

\address[3]{School of Mathematical Sciences, Rochester Institute of Technology, Rochester, NY 14623}

\cortext[cor1]{Corresponding author}

\nonumnote{In this work, we show that Discrete Wavelet Transform is the best method for feature extraction in vocal signals.}

\begin{abstract}
Facing the diversity and growth of the musical field nowadays, the search for precise songs becomes more
and more complex. The identity of the singer facilitates this search. 
In this project, we focus on the problem of identifying the singer by using different methods for feature extraction. Particularly, we introduce the Discrete Wavelet Transform (DWT) for this purpose. To the best of our knowledge, DWT has never been used this way before in the context of singer identification. This process consists of three crucial parts. First, the vocal signal is separated from the background music by using the Robust Principal Component Analysis (RPCA). Second, features from the obtained vocal signal are extracted. Here, the goal is to study the performance of the Discrete Wavelet Transform (DWT) in comparison to the Mel Frequency Cepstral Coefficient (MFCC) which is the most used technique in audio signals. Finally, we proceed with the identification of the singer where two methods have experimented: the Support Vector Machine (SVM), and the Gaussian Mixture Model (GMM). We conclude that, for a dataset of 4 singers and 200 songs, the best identification system consists of the DWT (db4) feature extraction introduced in this work combined with a linear support vector machine for identification resulting in a mean accuracy of 83.96\%.\\
\end{abstract}

\maketitle

\section{Introduction}
Music is a universal art form and cultural activity which can have several effects on the listener depending on the intention of the artist as well on the state of mind of the listener. Hence, with music, it is possible to express critics against politics or society, mobilize people for a course or to point out feelings arising from love, happiness, sadness, or loneliness. With the increasing possibilities to access and to share art, the world of music is becoming more and more vast and diverse. Query fast through this world and collecting precise information is a challenge data scientists face today. In this sense, by listening to a song, one could develop interest in the biography of the artist and may want to access other songs from this artist. This issue on which this project is based on is known as the identification of the singer. The identification of the singer is done in three phases: the separation of the singer’s voice from the background music, the feature extraction and the identification process using the features extracted from the vocal signal obtained from the separation procedure.

\section{Background}
A great deal of research has been done in the field of singer identification. In 2002, \citep{liu2002singer} proposed a singer identification technique for the classification of MP3 musical objects according to their content. They used phoneme segmentation for signal separation. Unfortunately, the signal of the singer’s voice at the output of this method still contains a lot of background music (noise) which make the singer identification difficult.\\
In 2004, a spectrum-based method of identifying the singer, proposed by \citep{bartsch2004singing}, worked well only for ideal cases that contained audio samples with the singer’s voice only.
The test set accuracy was 70-80\%. In the ”Identification of the singer based on vocal and instrumental
models” proposed by  \citep{maddage2004singer}. in the same year, the singer was identified using both low-level characteristics and knowledge of musical structure. Using the dataset with 100 popular songs of solo singers, they obtained an accuracy of over 87\%. However, this method was not suitable for music that was more instrumental than singing.\\
A systematic approach to identify and separate the unvoiced singing voice from the musical accompaniment is proposed by \citep{hsu2009improvement}. For the separation of the singer’s voice, they used the spectral
subtraction method. This method follows the framework of Computer Auditory Scene Analysis (CASA)
which includes the segmentation and clustering steps. This method considerably improved the clarity of the singing voice signal but was not always sufficient because, during clustering, a lot of information is lost.\\
To solve the problem of identifying the singer based on the acoustic variables of the singer’s voice,
\citep{yang2016statistical} used the Gaussian Mixture Method (GMM) and Support Vector Machine (SVM) in 2016. He
obtained accuracies of 96.42\% and 81.23\% with a dataset of hundred (100) songs of ten (10) singers. For signal separation, he used Robust Principal Component Analysis (RPCA) which is an improved version of Principal Component Analysis (PCA) and gives a better result than NMF. For feature extraction, he used Mel-Frequency Cepstral Coefficient (MFCC).\\
In 2017, \citep{xing2017singer} proposed an effective system of singer identification with human voice separated from original music. He used first, Robust Principal Component Analysis (RPCA) to music separation with its high performance. After the clear enough human voices are extracted, the Linear Predictive Coding (LPC) method was chosen as the experimental method for feature extraction. Finally, the singer would be identified by Gaussian Mixture Model (GMM) with 63.6\% of accuracy in a dataset of 100 singers.\\
In 2019, The work of \citep{nameirakpam2019singer} had implemented discret wavelet transform (DWT) as a pre-processing step (denoising) prior to feature extraction to investigate the performance of singer's identification with and without DWT. It is found that after applying wavelet transform the accuracy result decreases. However, the decrease in percentage accuracy is minimal (5.79\%, 0.72\% and 0.72\% for 8, 16 and 32 Gaussians respectively). While the computational time is drastically reduced.

\section{Problem Statement and Contribution of this Study}
The recent and improved research presented above shows that singer's identification using DWT for pre-processing and MFCC for feature extraction is done in a much reduced time, but the accuracy decreases compared to the results obtained without DWT. Unfortunately, MFCC uses the Fast Fourier Transform (FFT) for the change from the time domain to the frequency domain. FFT does not retain the time domain information and results in loss of data during the change. In this study, we will use DWT for all the feature extraction process to see if it improves feature extraction more than MFCC because this method retains time-domain information by its ability to operate in both in the time and frequency domain simultaneously. The Robust Principal Component Analysis (RPCA) will be used to separate the singer's voice from the background music. To identify the singer, we will apply both the Support Vector Machine (SVM) and the Gaussian Mixture Model (GMM).
\section{Study Organization}
The objective of this study is to build a model allowing the identification of the speaker using DWT for feature extraction. To achieve this goal, we present Robust Principal Component Analysis (RPCA) as the best technique for separating the singer's voice and its methodology, followed by the description and process of feature extraction using Discrete Wavelet Transform (DWT). Then, we explain the learning techniques such as the Support Vector Machine (SVM) and the Gaussian Mixture Model (GMM) for singer identification. Finally, we present the experiments and results. We conclude our research and propose recommendations for future work.

\section{Singing Voice Separation Technique: RPCA}
\begin{figure}[h]
	\begin{subfigure}{0.3\textwidth}
		\includegraphics[width=1\textwidth]{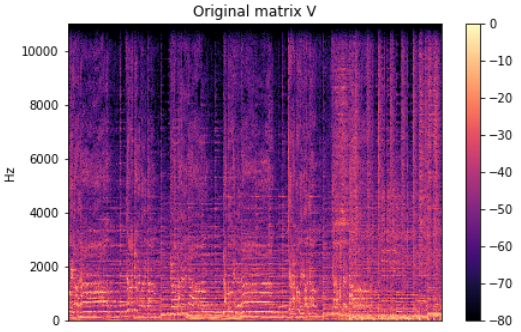}
		\caption{Original matrix V}
	\end{subfigure}
	\begin{subfigure}{0.3\textwidth}
		\includegraphics[width=1.15\textwidth]{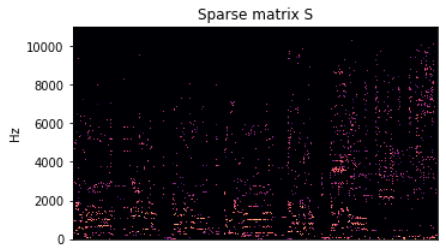}
		\caption{Sparse matrix S}
	\end{subfigure}
	\begin{subfigure}{0.35\textwidth}
		\includegraphics[width=1\textwidth]{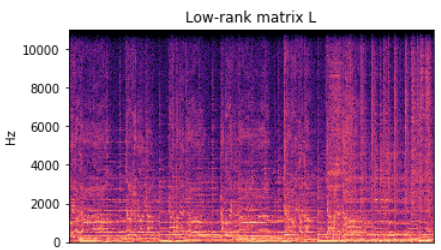}
		\caption{Low-rank L}
	\end{subfigure}
	\caption{Example RPCA results for Garou2.mp3: (a) the
		original matrix, (b) the low-rank matrix, and (c) the sparse matrix.}
	\label{rpca}
\end{figure}

Propsed by \citep{candes2011robust}, Robust Principal Component Analysis (RPCA) is a modification of the Principal Component Analysis (PCA) method. RPCA has been proven to perform well for noise-corrupted data compared to PCA. The idea of RPCA is to decompose a $V (V\in \mathbb{R}^{n\times p}$) data matrix into two other matrices, $L$ and $S$, as follows:
\begin{equation}
V=L+S
\end{equation}
In the case of sound, $L (L\in \mathbb{R}^{n\times p}$) is the low-rank matrix corresponding to the background music and $S (S\in \mathbb{R}^{n\times p}$) is the sparse matrix characterizing the singing voice. Indeed, in music, the part of noise incorporated (background music) often varies more slowly with time compared to the singer's voice. In other words, the singer's own voice is likely to be more non-stationary than the noise. This phenomenon can be easily observed by analyzing the spectrographs in Figure ~\ref{rpca}. The spectral structure of pure noise is usually fixed or slowly varying, while the vocal part changes rapidly over time. This hypothesis implies that the noise part appears to be of low-rank, while the pure voice part is sparse \citep{hung2018employing}. Therefore, extracting the sparse component from the music signal matrix tends to separate the background music from the voice of the speaker. This separation is made by convex optimization which can be written as \citep{candes2011robust}:
\begin{eqnarray}
\mbox{minimize}\ \lVert L\rVert_*+\lambda\lVert S\rVert_1 \label{eq32} \\ 
\mbox{subject to}\ L+S=V, \nonumber
\end{eqnarray}
where $\lambda>0$ is a trade-off parameter between the rank of $L$ and the sparsity of $S$, $\lVert.\rVert_*$  is the nuclear norm representing the sum of singular values of matrix entries and $\lVert.\rVert_1$ is the $L_1$-norm representing the sum of absolute values of matrix entries \citep{candes2011robust}. 

To solve the problem given to the equation \ref{eq32}, we use the Augmented Lagrange Multiplier (ALM) method. The corresponding formula is given by \citep{candes2011robust}:
\begin{equation}
L_a(L,S,Y,\mu)=\lVert L\rVert_*+\lambda \lVert S\rVert_1 +\langle Y,V-L-S\rangle+\frac{\mu}{2}\lVert V-L-S\rVert_F^2 \label{eq33}
\end{equation}
In equation \ref{eq33}, $\mu$ is a penalty parameter (always positive), $Y$ is slack variable matrix, $\lVert.\rVert_F$ is the Frobenius norm. $\langle Y,V-L-S\rangle$ implies the
standard trace inproduct. At the end, we obtain two matrices: the low-rank matrix $L$ and the sparse matrix $S$ respectively \citep{candes2011robust}.

\begin{figure}[h]
	\begin{subfigure}{0.3\textwidth}
		\includegraphics[width=1.1\textwidth]{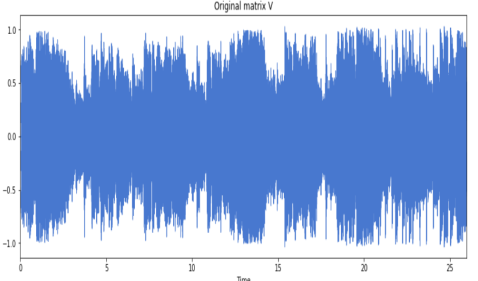}
		\caption{Original matrix V}
	\end{subfigure}
	\begin{subfigure}{0.3\textwidth}
		\includegraphics[width=1\textwidth]{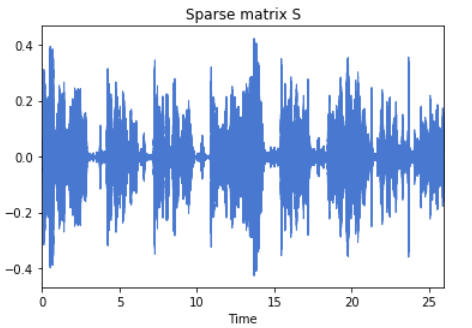}
		\caption{Sparse matrix S}
	\end{subfigure}
	\begin{subfigure}{0.35\textwidth}
		\includegraphics[width=0.8\textwidth]{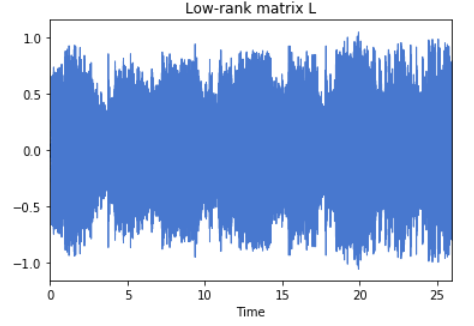}
		\caption{Low-rank L}
	\end{subfigure}
	\caption{Example of signal after ISTFT for Garou2.mp3: (a) the
		original matrix, (b) the low-rank matrix, and (c) the sparse matrix.}
	\label{istft}
\end{figure}

\begin{figure}[h]
	\centering
	\includegraphics[width=0.75\textwidth]{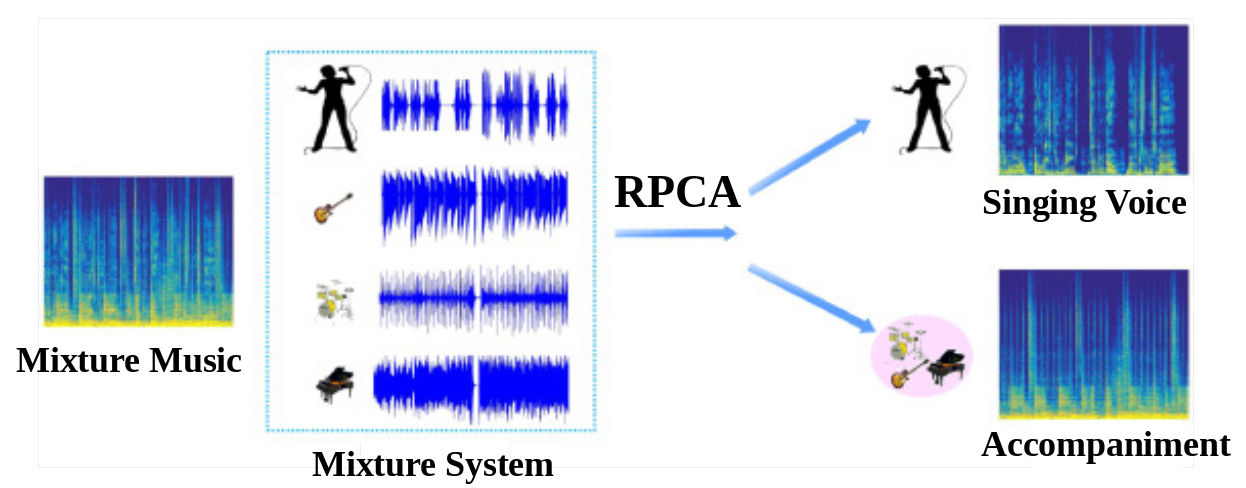}
	\caption{Signal separation process \citep{li2019blind}}
	\label{sep}
\end{figure}

To obtain the signals of the background music and the singing voice represented in Figure ~\ref{istft}, the Inverse Short-Time Fourier Transform (ISTFT) is performed to return to the temporal domain.
The signal separation process is summarized in Figure \ref{sep}.

\section{Discret Wavelet Transform (DWT) as a feature extraction technique}
Extraction of acoustic characteristics plays an essential role in the construction of a singer identification system. The objective is to select variables that have a high inter-label range and low discrimination power within the label. The discriminating power of characteristics or sets of characteristics indicates the extent to which they can discriminate between labels. The selection of characteristics is usually done by examining the discriminative power of the variables. The performance of a set of features depends on demand. Thus, designing them for a specific application is the main challenge in building singer identification systems.
In this section, we will present the theoretical background of the Discrete Wavelet Transform (DWT) method. DWT is based on dividing the signal into several sub-bands before performing feature extraction.

\subsection{Introduction of Wavelet Transform (WT)}
\subsubsection{Definition}~

Wavelet Transform (WT) is a very powerful tool for the analysis and classification of time series signals. It is unfortunately not known or popular in the field of data science. This is partly because you need to have some prior knowledge about signal processing, Fourier Transform and a solid mathematics background before you can understand the theory underlying the wavelet transform. However, we believe that it is also due to the fact that most books, articles and papers are far too theoretical and do not provide enough practical information illustrating how they could be used. 

WT has many applications in the analysis of stationary and non-stationary signals. These applications include removing noise from signals, detecting abrupt discontinuities, and compressing large amounts of data \citep{wang2013determination}. 
\subsubsection{Principle of WT}~

WT decomposes a signal into a group of constituent signals, called wavelets, each having a well-defined dominant frequency, similar to the Fourier Transform (FT) in which the representation of a signal is made by sine and cosine functions of unlimited duration. In WT, wavelets are transient functions of short duration, i.e. of limited duration centered around a specific time. The drawback of FT is that, as the time domain transitions to the frequency domain, information about what is happening in the time domain is lost. From the observation of the frequency spectrum obtained using FT, it is easy to distinguish the frequency content of the analyzed signal, but it is not possible to deduce in what time the signal components of the frequency spectrum will appear or disappear. Unlike FT, WT allows both time-domain and frequency-domain analysis, providing information on the evolution of the frequency content of a signal over time \citep{montejo2007aplicaciones}. There are many families of WT but the two principal are:
\begin{itemize}
	\item \textbf{Continuous Wavelet Transform (CWT)}: The values of the scaling and translation factors are continuous, which means that there can be an infinite amount of wavelets. It performs  a  multi-resolution  analysis by  contraction and  dilatation of  the wavelet functions \citep{aggarwal2011noise}. Its different sub-families are:
	\begin{itemize}
		\item Mexican hat wavelet
		\item Morlet wavelet
		\item Complex Gaussian wavelets
		\item Gaussian 
	\end{itemize}
	\item \textbf{Discrete Wavelet Transformations (DWT)}: It uses filter banks  for the construction of the  multi-resolution time-frequency plane and  special wavelet filters for the analysis and reconstruction of signals \citep{merry2005wavelet}. Its different sub-families are:
	\begin{itemize}
		\item Daubechies
		\item Symlets
		\item Coiflets
		\item Biorthogonal
	\end{itemize}
\end{itemize}
\subsection{Discrete Wavelet Transformations (DWT)}
DWT is defined by the following equation \ref{00}:
\begin{eqnarray}\label{00}
W ( j , k ) = \sum_n x ( n ) 2^{-\frac{j}{2}}\psi( 2^{-j} n - k ),
\end{eqnarray}
where $\psi(t)$ is a time function with finite energy and fast decay called the mother wavelet. $W ( j , k )$ represents the wavelet coefficients, where $k$ denotes location, and $j$ denotes level.

DWT has four families: (1) Daubechies; (2) Symlets; (3) Coiflets; (4) Biorthogonal. Each type has a different shape, smoothness, and compactness and is useful for a different application. Since a wavelet has to satisfy only two mathematical conditions which are the so-called normalization and orthogonalization constraints, it is easy to generate a new type of wavelet. 

DWT contains three major steps:
\begin{enumerate}
	\item \textbf{Wavelet Threshold De-Noising}
	
	In general, after separation of the voice signal, this signal still contains some small noises. The elimination of this noise is very important for the accuracy of the characteristics that will be extracted from the signal. Indeed, a singer has many different sounds and therefore, if the vocal signal extracted from these sounds contains noise, the identification will not be optimal. Donoho has introduced the use of wavelets to denoise the signals. He developed linear denoising for noises composed of high-frequency components and non-linear denoising (wavelet shrinkage) for noises also existing in the low frequencies \citep{donoho1995noising}. Schremmer et al. have developed software for real-time wavelet noise canceling of audio signals. Noise suppression is achieved by using soft or hard thresholding of the DWT of the coefficients \citep{schremmer2001wavelet}. The success criterion for noise suppression is the difference between the original signal and the denoised signal. A new speech enhancement system based on a wavelet denoising framework has been introduced by Fu Qiang and Wan Eric. In this system, noisy speech is first pre-processed using a generalized spectral subtraction method to initially reduce the noise level with negligible speech distortion. Then, the decomposition of the resulting speech signal into critical bands is done using the perceptual wavelet \citep{fu2003perceptual}. Denoising using DWT is developed in \citep{saric2005white} where the threshold is given by the equation:
	\begin{eqnarray}\label{tresh}
	\lambda=\sigma_n\sqrt{(2\log N)},
	\end{eqnarray}
	where $\lambda$ is the wavelet threshold, $\sigma_n$ is the standard deviation of the noise, and $N$ is the length of the sample signal.
	
	\item \textbf{Wavelet Decomposition}
	
	As illustrated in Figure \ref{dwt}, DWT breaks down a signal into several scales representing different frequency bands. Short-duration wavelets are used to extract information from the high-frequency components. Long-duration wavelets can be used to extract information from low frequencies  \citep{chang2000adaptive}. The process goes on under multiple levels as a subsequent coefficient from the first level within the approximation. At each process, the frequency resolution is doubled using the filters while decomposing and reducing the time complexity to half. In the end, we consider all the high-frequency bands ($H_1, H_2, H_3, H_4$) and the last low-frequency band ($L_4$).
	\begin{figure}[h]
		\centering
		\includegraphics[width=0.9\textwidth]{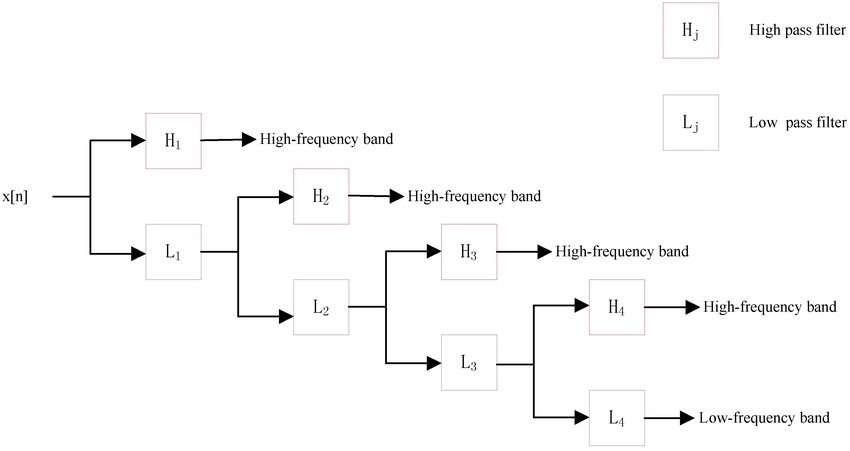}
		\caption{DWT process level four \citep{dwt})}
		\label{dwt}
	\end{figure}
	
	\item \textbf{Feature Extraction}
	
	Multi-resolution analysis (MRA) is used to extract feature vectors from the signal data. Very common in vocal signal, time-frequency domain DWT  based statistical features for classification include mean average value, standard deviation, and spectral entropy.
	\begin{itemize}
		\item \textbf{Mean average value:} it defines the mean of each vector of the sub-bands obtained in the previous step. It is given as,
		\begin{eqnarray}
		\mu=\frac{\sum_{i=1}^Nx_i}{N}
		\end{eqnarray}
		
		\item \textbf{Standard deviation:} it defines the variance of the signal. It is given by
		\begin{eqnarray}
		\sigma=\dfrac{\sum_{i=1}^{N}(x_i-\mu)^2}{N}
		\end{eqnarray}
		
		\item \textbf{Power spectral density:} it is calculated in two steps: first, by finding the Fast Fourier Transform (FFT) $F(\xi_i)$ of the time series and then, taking the squared modulus of the FFT coefficients.
		\begin{eqnarray}
		P(\xi_i)=\frac{|F(\xi_i)|^2}{N}
		\end{eqnarray}
		\item \textbf{Spectral entropy:} it is the measure of randomness and the information content of a signal. To calculate the entropy of a given vocal signal, we use the Shannon entropy formula
		\begin{eqnarray}
		E=-\sum_{i=1}^{N}x_i^2\log(x_i^2)
		\end{eqnarray}
	\end{itemize}
\end{enumerate}
\section{Learning Techniques: Classification Models}
In the world of machine learning, two main areas can be distinguished: supervised learning and unsupervised learning. The main difference between the two lies in the nature of the data and the approaches used to process them. In this section, we present two learning techniques that are widely used for audio: Support Vector Machine (SVM) and Gaussian Mixture Model (GMM).

\subsection{Support Vector Machine (SVM)}
Support Vector Machine (SVM) was developed by Cortes and Vapnik in 1995 and improved by Boser, Guyon, and Vapnik in 1998 \citep{boser1992training}; \citep{vapnik1998statistical} which is useful for solving problems of monitoring classification in high dimensions. The SVM approach searches directly for a plane or surface of separation by an optimization procedure that finds the points that form the boundaries of the classes. These points are called support vectors. Besides, the SVM approach uses the kernel method to map the data with a nonlinear transformation to a high-dimensional space and tries to find a separation surface between the two classes in this new space. When we have two labels (classes), we use the binary SVM and in cases with more than two labels, we apply the multi-SVM.
\subsubsection{Binary SVM}~

\begin{figure}[h]
	\centering
	\includegraphics[width=0.4\textwidth]{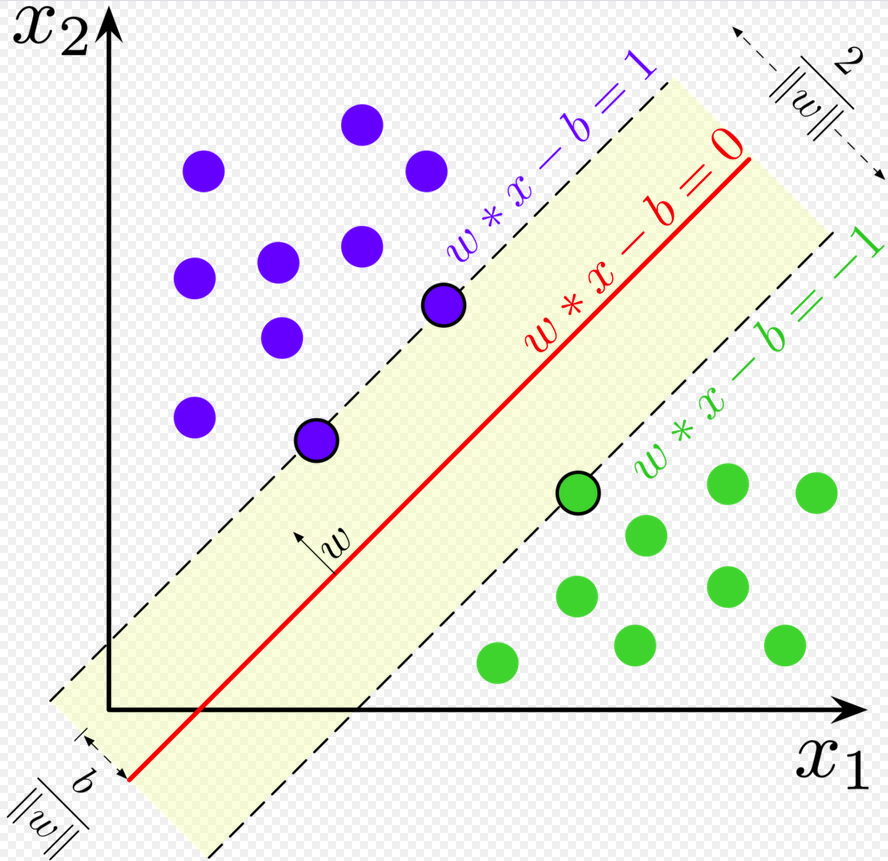}
	\caption{SVM graph \citep{svm}})
	\label{svm}
\end{figure}

Binary-SVM is used when the data has exactly two classes. For classification, SVM finds the best hyperplane that separates all data points of one class from those of the other class as illustrated in Figure \ref{svm} by the red line. The best hyperplane is the one with the largest margin between the classes. 

The hyperplane equation is given by
\begin{eqnarray}
w^Tx - b = 0,
\end{eqnarray}
where $w$ is the weight and $b$ the bias. $x$ is an input variable. 

\begin{figure}[htbp!]
\centering
\includegraphics[width=0.5\textwidth]{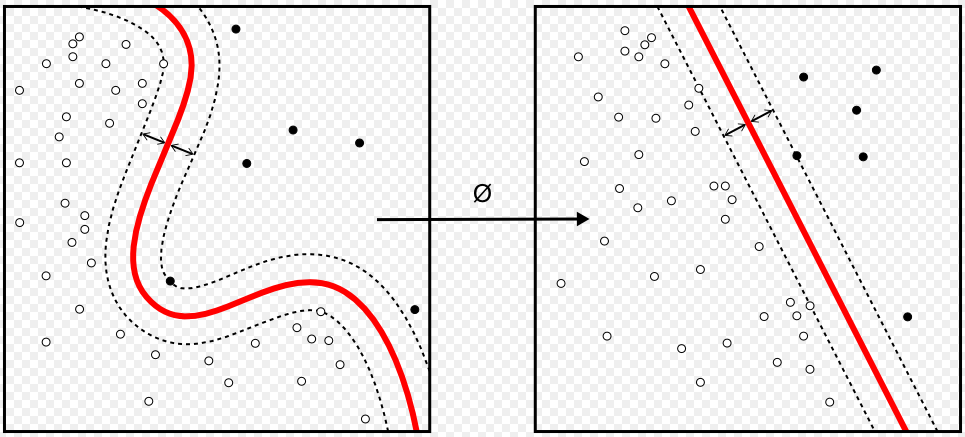}
\caption{Kernel SVM graph \citep{svm}}
\label{ker}
\end{figure}

There are three cases of Binary-SVM:
\begin{enumerate}
	\item \textbf{Hard margin}
	
	Here, the two classes can be separated linearly (Figure \ref{svm}). The goal is to maximize $\frac{2}{\lVert w\rVert}$ which is equivalent to minimizing $\lVert w \rVert$. 
	
	Hence, the problem can be reformulated to $w^* = argmin\ \lVert w \rVert^2$.
	We have two constraints:
	\begin{enumerate}
		\item $	w^T x^{(i)} - b \geq 1,\ if\ y^{(i)} = 1$
		\item $w^T x^{(i)} - b \leq -1,\ if\ y^{(i)} = -1$
	\end{enumerate}	
	We combine these two constraints and get
	\begin{eqnarray*}
		y^{(i)} (w^T x^{(i)} - b) \geq 1,\qquad for\ i = 1, . . . , m,
	\end{eqnarray*}
	where $m$ is the number of samples and $y$ is the label.
	
	So, the optimization problem becomes:
	\begin{eqnarray*}
		\begin{cases}
			w^* = \mbox{argmin}\ \lVert w \rVert^2\\
			\mbox{subject to}\ y^{(i)} (w^T x^{(i)} - b) \geq 1
		\end{cases}
	\end{eqnarray*}
	The SVM classifier is given by:
	\begin{eqnarray}
	f (\vec{x}) = \mbox{sign}( \vec{w}.\vec{x} - b)
	\end{eqnarray}
	\item \textbf{Soft margin}
	
	This case occurs when the data are not separable in a linear way because there are dots within the margin. Consequently, the loss of function becomes the hinge loss:
	\begin{eqnarray}\label{1}
	\ell(w; x, y) = \mbox{max}[0, 1 - y^{(i)} (w^T x^{(i)} - b)],\ i=1,...,m.\ 
	\end{eqnarray}
	
	Having the hinge loss by equation \ref{1}, the expected loss is given by
	\begin{eqnarray*}
		L(w)=\frac{1}{m}\left[\sum_{i=1}^m\ell(w; x, y)\right]+\lambda \lVert w \rVert_2^2,
	\end{eqnarray*}
	where $\lambda$ is the trade-off increasing the size of the margin and ensuring that the data point is on the correct side of the margin.
	
	Hence, the optimization problem becomes:
	\begin{eqnarray*}
		\begin{cases}
			w^* = \mbox{argmin}\ L(w)\\
			\mbox{subject to}\ y^{(i)} (w^T x^{(i)} - b) \geq 1- \ell
		\end{cases}
	\end{eqnarray*}
	\item \textbf{Nonlinear classification (kernel SVM)}

In Figure \ref{ker} a case where the data cannot be separated by a hyperplane is depicted. We find a map $\phi : x^{(i)} \longrightarrow \phi(x)^{(i)}$ from the data space to the feature space such that the data are linearly separable in the feature space by applying the so-called ``kernel trick":
\begin{eqnarray}\label{411}
k(x, x_i ) = \phi(x) \phi(x_i)
\end{eqnarray}
Kernel function may be any of the symmetric functions that satisfy the Mercer's conditions \citep{brunner2012pairwise}.

In the feature space, one can write:
\begin{eqnarray}\label{412}
w^T &=& \sum_{i=1}^n\alpha_iy^{(i)}\phi(x^{(i)})\nonumber \\
\Longrightarrow w^T \phi(x^{(i)})&=& \sum_{i=1}^n\alpha_iy^{(i)}\phi(x^{(i)})\phi(x^{(i)})
\end{eqnarray}
Using \ref{411} in \ref{412}, we obtain:
\begin{eqnarray*}
	w^T \phi(x^{(i)})= \sum_{i=1}^n\alpha_iy^{(i)}k(x^{(i)},x^{(i)} )
\end{eqnarray*}
So,
\begin{eqnarray*}
	y^{(i)} (w^T x^{(i)} - b) \Longrightarrow  y^{(i)} \left[\sum_{i=1}^n\alpha_iy^{(i)}k(x^{(i)},x^{(i)} ) - b\right]
\end{eqnarray*}
There are several functions of the SVM kernel.
\begin{enumerate}
	\item \textbf{Polynomial kernel:} it is a non-stationary kernel. The polynomial kernel is well suited for problems where all training data are normalized. It is given by  equation \ref{413}
	\begin{eqnarray}\label{413}
	K (x, x_i ) = ( \alpha x^T x_i + c)^d,
	\end{eqnarray}
	where the slope $\alpha$ is the adjustable parameter, $d$ is the polynomial degree and $c$ is the constant.
	
	The dimension of the feature space vector $\phi(x)$ for the polynomial kernel of degree $p$ and for the input pattern of dimension $d$ is:
	\begin{eqnarray*}
		\frac{(p+d)!}{p!d!}
	\end{eqnarray*}
	\item \textbf{Gaussian kernel:} it is an example of radial basis function kernel (RBF). It is characterized by the equation: 
	\begin{eqnarray}\label{414}
	k(x, x' ) = exp[-\gamma \lVert x - x'\rVert^ 2 ]
	\end{eqnarray}
	Usually, $\gamma=\frac{1}{2\sigma^2}$. So, the equation \ref{414} becomes:
	\begin{eqnarray}
k(x, x' ) = exp\left[-\frac{\lVert x - x'\rVert^ 2}{2\sigma^2}\right] 
\end{eqnarray}
The adjustable parameter $\sigma$ plays a major role in the performance of the kernel and should be carefully tuned to the problem at hand. If overestimated, the exponential will behave almost linearly and the high-dimensional projection will begin to lose its non-linear power. On the contrary, if underestimated, the function will lack regularization and the decision boundary will be highly sensitive to noise in training data \citep{ramalingam2014speech}.

\end{enumerate}
\end{enumerate}

SVM classifier is given by:
\begin{eqnarray}
f (x) = \mbox{sign}(w^T\phi(x) + b)
\end{eqnarray}

\subsubsection{Multi-Class SVM}~

SVM was made for binary classification. But in the real world, we deal with classification problems with more than two classes. Multi-category classification problems are usually divided into a series of binary problems so that binary SVM can be directly applied \citep{mathur2008multiclass}. 

One representative method is the ``One-Against-All" approach. Consider an $M$-class problem, where we have $N$ training samples: $\{x^{(1)},y^{(1)}\},..., \{x^{(N)},y^{(N)}\}$. Here, $x^{(i)}\in \mathbb{R}^m$ is a $m$-dimensional feature vector and $y^{(i)}\in \{1,2,...,M\}$ is the corresponding class label.

The One-Against-All approach constructs $M$ binary SVM classifiers, each of which separates one class from all the rest. The $i^{th}$ SVM is trained with all the training examples of the $i^{th}$ class with positive labels and all the others with negative labels. 

Mathematically, the $i^{th}$ SVM solves the following problem that yields the $i^{th}$ decision function $f_i(x) = sign(w_i^T\phi(x) + b_i)$
\begin{eqnarray}
\mbox{minimize}&:&\qquad L(w,\xi_j^i)=\frac{1}{2}\lVert w_i\rVert^2 +C\sum_{l=1}^N\xi_j^i\nonumber\\
\mbox{subject to}&:&\qquad  \tilde{y}_j(w_i^T\phi(x_j)+b_i)\geq 1-\xi_j^i,\quad \xi_j^i\geq0,
\end{eqnarray}
where $\tilde{y}_j=1$ if $y_j=i$ and $\tilde{y}_j=-1$ otherwise.

At the classification phase, a sample $x$ is predicted to be in class $i^*$ whose $f_{i^*}$ produces the largest value
\begin{eqnarray}
i^*=\mbox{argmax} f_i(x)=\mbox{argmax}(w_i^T\phi(x) + b_i),\quad i=1,...,M
\end{eqnarray}

\subsection{Gaussian Mixture Model (GMM)}
Gaussian Mixture Model (GMM) is a parametric probability density function expressed as a weighted sum of Gaussian component densities. In a biometric system, GMMs are widely used as a parametric model of the probability distribution of continuous measurements or features, such as spectral features related to the vocal tract in a speaker recognition system. GMM parameters are estimated from training data using the iterative expectation maximization (EM) algorithm \citep{reynolds2009gaussian}.

\subsubsection{K-means clustering}~

Here, the initialization of the GMM parameters is carried out using the number of clusters and allowing to form the different centers.
In fact, GMM is a function that is comprised of several Gaussians, each identified by $k \in \{1,..., K\}$ ($K$: number of clusters of the dataset).

\begin{figure}[h]
	\centering\includegraphics[width=7cm]{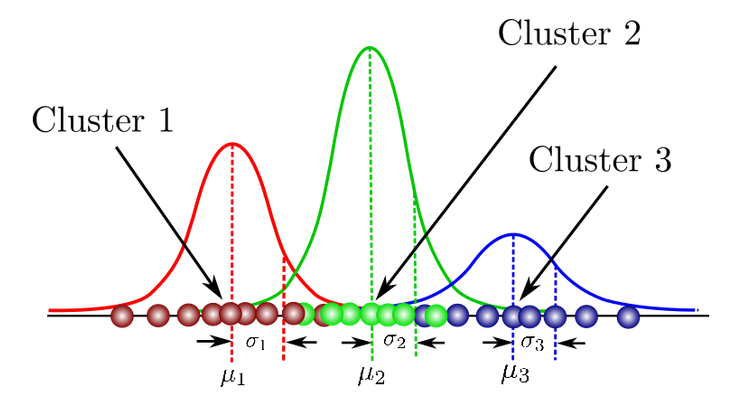}\
	\caption{GMM parameters ~(Reynolds et al., \citep{reynolds2009gaussian}).}
	\label{gmm}
\end{figure}

Each $k$ has the following parameters  (Figure \ref{gmm}):
\begin{enumerate}
	\item \textbf{A mean vector $\vec{\mu}$} that defines its center.
	\item \textbf{A covariance matrix $\Sigma$} that defines its width. This would be equivalent to the dimensions of an ellipsoid in a multivariate scenario.
	\item \textbf{A mixture of weights $w$} that defines how big or small the Gaussian function will be.
\end{enumerate}	

A mixture of weights must satisfy the constraint that:
\begin{eqnarray}
\sum_{k=1}^K w_k=1
\end{eqnarray}
GMM is giving by
\begin{eqnarray}
p(x|\lambda)=\sum_{k=1}^Kw_k\ g(x|\mu_k,\Sigma_k),\qquad k = 1, . . . , K,
\end{eqnarray}
where $x$ is the D-dimensional continuous-valued data vector (i.e. measurement or features), $w_k$ are the mixture weights, and $g(x|\mu_k,\Sigma_k)$ are the Gaussian densities. Each component density is a D-variate Gaussian function of the form:
\begin{eqnarray}
g(x|\mu_k,\Sigma_k)	=\frac{1}{(2\pi)^{D/2}|\Sigma_i|^{1/2}}	exp\left[-\frac{1}{2}(x-\mu_k)^T\Sigma^{-1}(x-\mu_k)\right]
\end{eqnarray}
A collective representation of the parameters is defined as:
\begin{eqnarray}\label{424}
\lambda=\{w_k,\mu_k,\Sigma_k\},\qquad k = 1, . . . , K	
\end{eqnarray}
There are several variants of the GMM presented in equation \ref{424}. The $\Sigma_k$ can be full rank or forced to be diagonal. Also, parameters can be shared or linked between Gaussian components. As an example, by having a common covariance matrix for all components, the choice of model configuration (the number of components, full or diagonal covariance and parameter coupling) is usually determined by the volume of data available for the estimation of GMM parameters, and by the way in which the GMM is used in a particular biometric application. It is important to note this because even if the characteristics are not statistically independent, the Gaussian components act together to model the overall density of the characteristics. The modeling of correlations between the vector components of the features can be performed by the linear combination of the Gaussian diagonal covariance basis. The effect of employing a set of $K$ full covariance matrices Gaussian can also be obtained by employing a larger set of Gaussian diagonal covariance matrices.

The use of a GMM to represent feature distributions can also be driven by the intuitive idea that the densities of individual components can model an underlying set of hidden classes. For example, in the case of the speaker, it is reasonable to assume that the acoustic space of the spectral features corresponding to a speaker's major phonetic events such as vowels or fricatives. These acoustic classes reflect certain general configurations of speaker-dependent vocal pathways that are useful in characterizing speaker identity. The spectral shape of the $k^{th}$ acoustic class can, in turn, be represented by the mean $\mu_k$ of the $k^{th}$ component density, and variations in the mean spectral shape can be represented by the covariance matrix $\Sigma_k$. Since not all the characteristics used to form the GMM are labeled, the acoustic classes are hidden, in the sense that the class of observation is unknown. The observation density of the feature vectors derived from these hidden acoustic classes form a Gaussian mixture (assuming that the feature vectors are independent) \citep{reynolds2009gaussian}.

\subsubsection{Expectation-Maximization (EM)}~

Taking into account the training vectors and a configuration of the GMM, we wish to estimate the parameters of the GMM, $\lambda$, which in some sense corresponds best to the distribution of the training vectors. There are several techniques for estimating the parameters of a GMM \citep{mclachlan1988mixture}. By far, the most popular and best-established method is the expectation-maximization (EM) algorithm. The objective of expectation maximization (EM) is to find the model parameters that maximize the probability of the GMM given the training data. For a sequence of $T$ training vectors $X = {x_1, ... . x_T}$, the likelihood of the GMM, assuming independence between the vectors, can be written as 
\begin{eqnarray}
p(\vec{x}|\Theta)=\sum_{k=1}^Kw_kp_k(\vec{x}|z_k,\theta_k),
\end{eqnarray}
where $p(\vec{x})$ is the finite mixture model, and $p_k(\vec{x}|z_k,\theta_k)$ is the gaussian density for the $k^{th}$ mixture component. $z = (z_1 , ..., z_K )$ is the vector of $K$ binary indicator variables which are mutually exclusive and exhaustive. $\Theta$ is the complete set of parameters ($\Theta=\{w_1,...,w_K,\theta_1,...,\theta_K\}$)

\textbf{Step 1: E-step }

The goal here is to compute the membership weights $w_{ik}$ which are the probabilities that reflect the uncertainty, given $\vec{x _i}$ and $\Theta$. The membership weight of a data point $\vec{x _i}$ in cluster $k$ can be written as: 
\begin{eqnarray}\label{426}
w_{ik}=p(z_{ik}=1|\vec{x}_i,\Theta)=\frac{w_kp_k(\vec{x}|z_k,\theta_k)}{\sum_{m=1}^Kw_mp_m(\vec{x}_i|z_m,\theta_m)}
\end{eqnarray}

\textbf{Step 2: M-step}~

This step aims to use the membership weights obtained in equation \ref{426} in E-step, to calculate new parameter values which are given by equation \ref{427}, \ref{010}, \ref{000}.
\begin{eqnarray}\label{427}
w_k&=&\frac{N_k}{N},\ 1\leq k\leq K,
\end{eqnarray}
\begin{eqnarray}\label{010}
\vec{\mu}_k&=&\frac{1}{N_k}\sum_{i=1}^Nw_{ik}\vec{x}_i,\ 1\leq k\leq K
\end{eqnarray}
\begin{eqnarray}\label{000}
\Sigma_k&=&\frac{1}{N_k}\sum_{i=1}^Nw_{ik}(\vec{x}_i-\vec{\mu}_k)(\vec{x}_i-\vec{\mu}_k)^T,\ 1\leq k\leq K.
\end{eqnarray}
where $N_k=\sum_{i=1}^Nw_{ik}$ is the column sum of the membership weight matrix.

\section{Implementation and Analysis}
This section presents the implementation of the techniques studied in the previous sections for the singer's identification. We will first give an overview of the procedure from separation to identification. Then, we will present all the experimental data used in this study. Finally, we will show the experiments and results obtained at each step of the singer identification process with and without feature extraction. 
\subsection{The Block Diagram of the Overall Process of Singer's Identification}
\begin{figure}[h]
	\centering
	\includegraphics[width=0.9\textwidth]{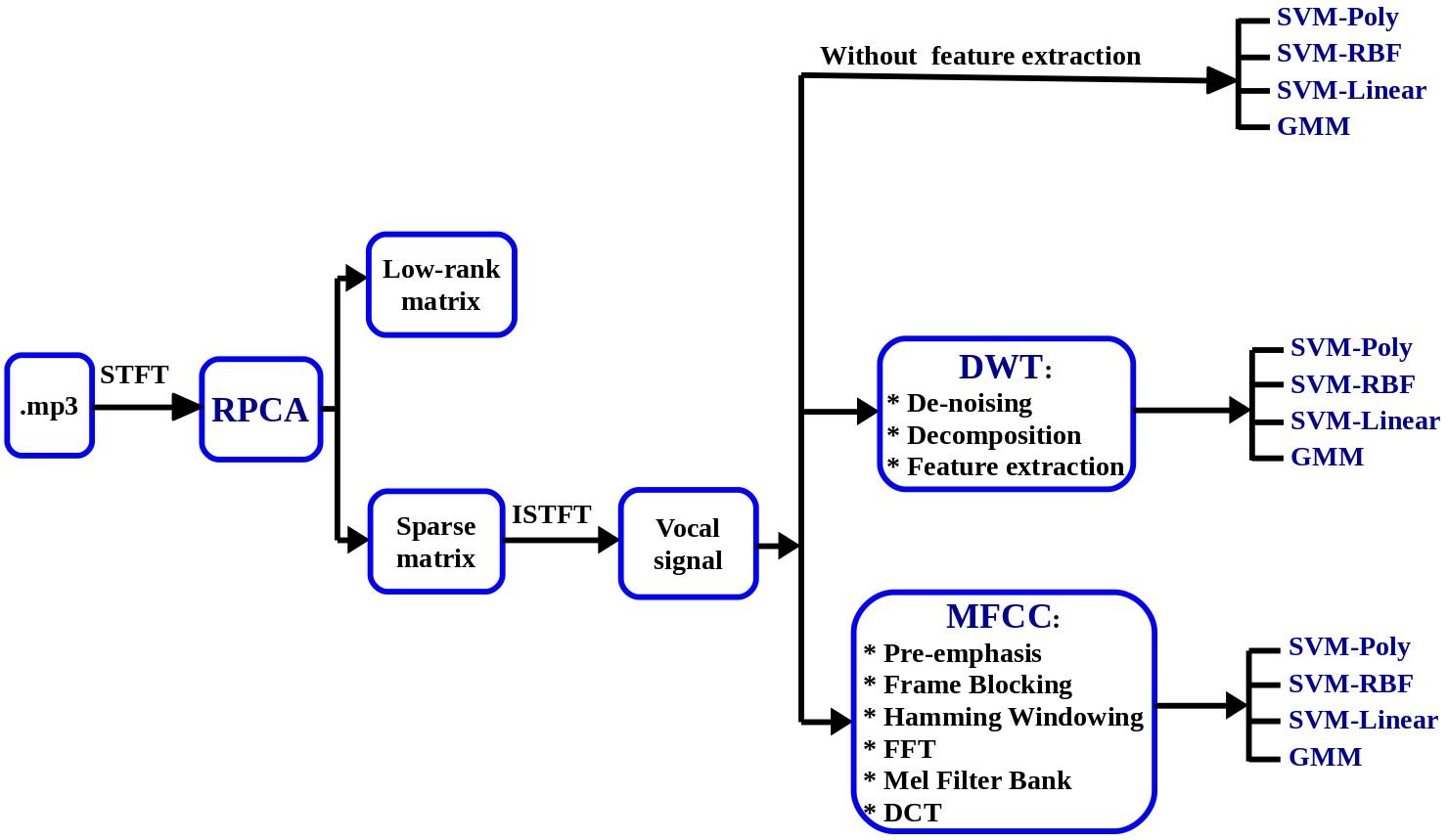}
	\caption{Block diagram of the overall process of singer's identification}
	\label{blo}
\end{figure}

As shown in the Figure \ref{blo}, the inputs are audio files with the .mp3 extension. After importing these files, we get the musical signal. We apply the STFT to each signal to obtain its matrix in the frequency domain. Afterward, the RPCA technique is applied to this matrix separating it into two matrices: a low rank and a sparse matrix. After performing the ISTFT on the sparse matrix, the vocal signal is obtained. This vocal signal obtained from each sound of the dataset allows building a data-frame. The purpose of this study is to show, first, the importance of feature extraction and then, to compare the two techniques DWT and MFCC. Hence, we perform three experiments: (1) Training the data without feature extraction; (2) Using MFCC for feature extraction; (3) Using DWT for feature extraction before training with SVM and GMM techniques.

\subsection{Dataset}
We created a database of test recordings by selecting four popular singers: two men and two women, each with 50 excerpts. These songs go through the pre-processing phase where missing values are removed and twelve-second singing voice segments are obtained from each musical recording. As a result, after separation of signal and after using RPCA, each singer has a total of 263232 singing voice segments which are then introduced into the feature extraction phase.
\subsection{Feature extraction using DWT}
\subsubsection{De-noising}

\begin{figure}[h]
	\centering
	\includegraphics[width=1\textwidth]{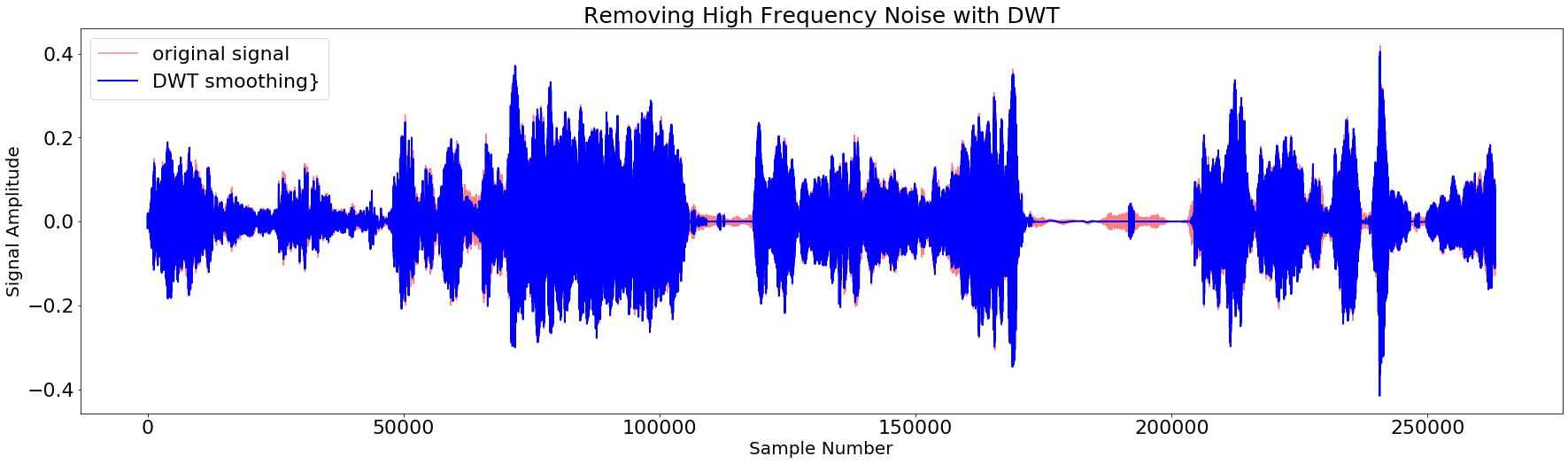}
	\caption{De-noising of the signal of Celine's song}
	\label{de}
\end{figure}

For de-noising, we investigate the maximum gain for input signals with different levels of degradation. The amount of white noise added to the original signal is controlled with the standard deviation of the noise $\sigma_n$. The maximum gain is obtained by
replacing the threshold $\lambda$ of equation \ref{tresh} with
\begin{eqnarray}\label{tre}
\lambda=k.\sigma_n\sqrt{(2\log N)},
\end{eqnarray}
where $0<k<1$. In fact, the universal threshold t given by equation \ref{tresh} is too high for audio signals and it cuts a big part of the original signal. So it is modified with factor $k$ to obtain a higher quality output signal. The value of $k$ is changed gradually with steps of 0.1. Finally, we find $k$ that gives the best result depicted in Figure \ref{de}.
\subsubsection{Decomposition of signal}

\begin{figure}[h]
	\centering
	\includegraphics[width=0.65\textwidth]{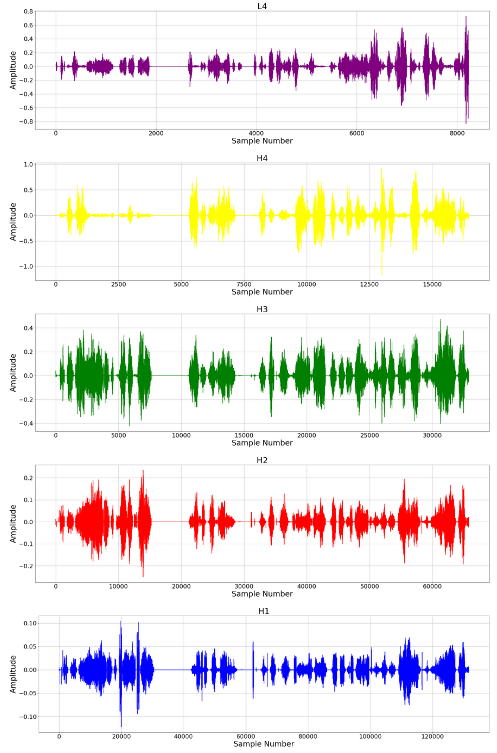}
	\caption{Decomposition of the signal of Celine's song previously de-noised}
	\label{decomp}
\end{figure}

According to literature, we use, for this implementation, the DWT Daubechies four (db4) of level 4. Each signal has the frequency $4160\ Hz$ after de-noising. In Figure \ref{decomp}, we can easily see the five (05) sub-bands of the previously de-noised song: $L_4 (0-260 Hz)$, $H_4 (260-520 Hz)$, $H_3 (520-1040 Hz)$, $H_2 (1040-2080 Hz)$, $H_1 (2080-4160 Hz)$
\subsubsection{Feature extraction strictly speaking}

After the decomposition of the signal in sub-bands, we extract the following features to build our final data-frame:
\begin{itemize}
	\item \textbf{Time-Frequency domain:} Mean and Spectral entropy.
	\item \textbf{Time domain:} Mean, Median, and Standard deviation.
	\item \textbf{Frequency domain:} Power spectral density.
\end{itemize}
\subsection{Singer identification}
In this step, we first do a feature engineering to see which of the 18 features are correlated and use PCA to take the features which represent 99.99\% of variation: 15 features. Then, we separate our data-frame: features for input and names of singers for output labels. We then check the parameter of models which can influence the result and take the one which gives us the best result. Finally, knowing the audio signal, we train the machine learning models using the fold cross-validation value 10. We also shuffle the data fifteen (15) times to have realistic results. Then, the general accuracy will be the mean of the fifteen (15) accuracies.\\

First, we do the training without feature extraction. We remark that the best model is delivered by the SVM-RBF model with 36.78\% of mean accuracy. It should be noted that the 263232 columns in the data-frame, without feature extraction, are considered as input for model training. Then, it becomes impossible to build the covariance matrix for GMM because the number of inputs is greater than the number of observations. However, a reduction in size is not possible because these columns represent the vector of the signal, and reducing it is equivalent to muting the signal, therefore, the implementation of this method was not possible.\\
Second, we do the training using MFCC for feature extraction. We found that, in general, the SVM model performs better than the GMM. The best model is SVM-Linear with an mean accuracy of 61.49\%.\\

Finally, we train the models using the final data-frame obtained at the previous step with DWT for feature extraction. We have determined that the best model is SVM-Linear with 83.96\%.\\

\begin{figure}[h]
	\centering
	\includegraphics[width=0.8\textwidth]{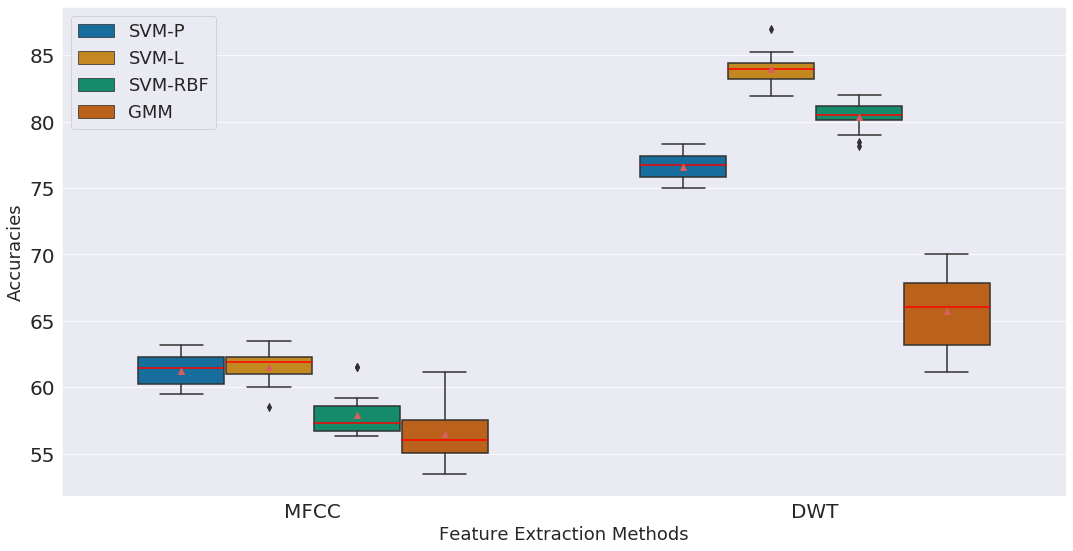}
	\caption{Boxplot of performance (accuracies) with DWT}
	\label{sum}
\end{figure}

The results have been summary in Figure \ref{sum}
\section{Conclusion and Recommendation}
The objective of this work was to apply the DWT for feature extraction and compare the results with the MFCC to see which of the two improves the identification of the singer in term of accuracy. We first gave the physical and mathematical description of the different techniques ranging from separation of the vocal signal from the background signal to the singer identification. Then, we implemented these techniques according to a dataset of 200 songs (50 songs per singer). RPCA was used for the separation of signals; DWT and MFCC were used to feature extraction; and SVM and GMM were used for singer's identification. 

For a set of 200 observations of audio signals, this study shows that:
\begin{itemize}
	\item The vocal signals are very unstable because the same artist sings in different styles and uses different nuances that influence the results of the study using a small number of recordings. As a consequence, low accuracies are obtained.
	\item The feature extraction is essential for the singer identification process.
	\item DWT performs better than MFCC for feature extraction in term of accuracy and training time.
	\item SVM performs better than GMM for singer identification.
	\item The best configuration of techniques for singer identification is DWT + SVM-Linear with a mean accuracy of 83.96\% with a training time of 1 068 s.
\end{itemize}

However, to generalize these results, it is essential to  perform the same study on a much larger set of recordings.

Since DWT is better at extracting characteristics from audio signals, future work could be the in-depth study of the different families of DWT to investigate the effects of their individual properties and to use appropriate DWTs for different cases of datasets.

Besides, DWT can be used for extended periods to study other non-stationary signals such as those in the human body (electroencephalogram (EEG), electrocardiogram (ECG), electro-oculography (EOG)). This will allow abnormalities to be detected fairly quickly and diseases predicted and treated before complications arise.

\printcredits
\bibliographystyle{cas-model2-names}

\bibliography{ref}
\end{document}